
\magnification=1200
\def\tr#1{{\rm tr #1}}
\def\d{\partial}
\def\f#1#2{{\textstyle{#1\over #2}}}

\def\next{\hfil\break\noindent}
\def\R{{\bf R}}
\def\nnext{\vskip 10pt\noindent}
\font\title=cmbx12

{\title
\centerline{Global properties of locally spatially homogeneous}
\centerline{cosmological models with matter}}

\vskip 2cm\noindent
Alan D. Rendall\footnote* {Present address: Institut des Hautes Etudes
Scientifiques, 35 Route de Chartres, 91440 Bures sur Yvette, France}
\next
Max-Planck-Institut f\"ur Astrophysik
\next
Karl-Schwarzschild-Str. 1
\next
Postfach 1523
\next
85740 Garching
\next
Germany

\vskip 2cm\noindent
{\bf Abstract}

The existence and nature of singularities in locally spatially
homogeneous solutions of the Einstein equations coupled to various
phenomenological matter models is investigated. It is shown that,
under certain reasonable assumptions on the matter, there are no
singularities in an expanding phase of the evolution and that
unless the spacetime is empty a contracting phase always ends
in a singularity where at least one scalar invariant of the
curvature diverges uniformly. The class of matter models treated
includes perfect fluids, mixtures of non-interacting perfect
fluids and collisionless matter.

\vskip 3cm
\noindent
{\bf 1. Introduction}

When trying to understand the global properties of solutions
of the Einstein equations, a tractable starting point is
the study of solutions with high symmetry. Assuming that
there is a symmetry group with three-dimensional spacelike
orbits leads to the class of spatially homogeneous spacetimes.
More generally, it is possible to consider spacetimes which
are locally isometric to spatially homogeneous ones. For
these spacetimes the Einstein equations reduce to ordinary
differential equations, an enormous simplification. At
the same time these spacetimes are general enough to display
a variety of interesting dynamical behaviour.

The purpose of this paper is to study the global properties of
locally spatially homogeneous cosmological models which admit
a compact Cauchy surface. Attention will be confined
to certain phenomenological matter models. No attempt will be
made to determine the details of the evolution of these
spacetimes; the aim is rather to answer two fundamental
questions. The first is whether singularities can
occur in an expanding phase of the evolution of the
spacetime and the second is whether a singularity which occurs
in a contracting phase must be a curvature singularity. A
review of results on these questions in the case where the
matter is a perfect fluid can be found in [6].

The class of cosmological models to be considered will now
be defined. They all admit Cauchy surfaces and have a
matter content which is such that the Cauchy problem for
the Einstein equations coupled to the equations describing
the matter model is well-posed. Hence it is enough
to define the class of initial data which gives rise to them.
The initial data consists of a Riemannian metric $g_{ab}$, a
symmetric tensor $k_{ab}$ and some matter data, denoted
collectively by $F_0$, on a three-dimensional manifold $M$.

\vskip 10pt\noindent
{\bf Definition} An initial data set $(g_{ab},k_{ab},F_0)$ for
the Einstein-matter equations is called {\it locally
homogeneous} if the naturally associated data set on the
universal covering manifold $\tilde M$ is homogeneous i.e.
invariant under a transitive group action.

\vskip 10pt\noindent
The terminology is motivated by the following fact. A
Riemannian manifold $(M,g)$ is called locally homogeneous
if given any two points $x,y\in M$ there exists an isometry
$\phi$ of a neighbourhood of $x$ onto a neighbourhood of $y$
with $\phi(x)=y$. Singer's theorem [14] says that a complete
Riemannian manifold $M$ is locally homogeneous if and only if
its universal cover is homogeneous. It is left open here
whether an analogue of Singer's theorem holds for initial
data sets for the Einstein equations. The spacetimes considered
in the following will be Cauchy developments of locally
homogeneous initial data sets on compact manifolds. It is
convenient for many purposes to work on the universal
covering space $\tilde M$. In the notation no distinction
will be made between objects on $M$ and their pull-backs to
$\tilde M$.

The universal cover of the given spacetime may not be spatially
homogeneous but it can be extended to be spatially homogeneous.
(This follows from the fact that the equations describing the
evolution of the geometry and the matter fields on the universal
cover can be written in a form where there is no explicit dependence
on the spatial variables.) It will be assumed
that such an extension has been made. Then the universal cover
has a preferred foliation by orbits. Each leaf of this foliation
has constant mean curvature. There is an induced foliation of
the original spacetime by constant mean curvature hypersurfaces.
It is then topologically of the form $M\times I$, where $I$ is
an interval. A time coordinate can be defined by requiring that it be
zero on the initial hypersurface and that it be an arc length
parameter increasing towards the future on the timelike geodesics
normal to the initial hypersurface. This time coordinate is constant
on each leaf of the preferred foliation. A time coordinate of
this kind will be called Gaussian. Let the mean curvature of the
leaf with time coordinate $t$ be denoted by $H(t)$. In the
following the time orientation will be chosen so that the
mean curvature of the initial hypersurface $t=0$ is non-positive.
In other words the model is initially non-contracting.

The spatially homogeneous spacetimes can be divided into two
classes, the Bianchi models and the Kantowski-Sachs models.
(For general information on this see [16].) Most of this paper
is concerned with the Bianchi case. The Kantowski-Sachs case
is discussed in Section 5. For Bianchi models the universal
cover $\tilde M$ can be identified with a Lie group $G$.
Tensors on $G$ will be described in terms of frame components
in a left invariant frame. The components of the induced
metric and second fundamental form of the homogeneous hypersurfaces
will be denoted by $g_{ij}$ and $k_{ij}$ respectively. They are
functions of $t$. The mean curvature is given by $H=g^{ij}k_{ij}$.

The matter models to be considered will be defined in terms of some
general properties. As usual $T^{\alpha\beta}$ denotes the
energy-momentum tensor. When a specific matter model has been
chosen $T^{\alpha\beta}$ will be a functional of some matter
variables, denoted collectively by $F$, and the spacetime metric
$g_{\alpha\beta}$. In the following another quantity $N^\alpha$
(called the particle current density) will be required. It is
also assumed to be a functional of $F$ and $g_{\alpha\beta}$.
Now various properties which will be assumed at appropriate
points will be listed.

\nnext
(1) $T^{\alpha\beta}V_\alpha W_\beta\ge 0$ for all future-pointing
timelike vectors $V^\alpha$ and $W^\alpha$ (dominant energy
condition)

\nnext
(2) $T^{\alpha\beta}(g_{\alpha\beta}+V_\alpha V_\beta)\ge 0$ for
all unit timelike vectors $V^\alpha$ (non-negative sum pressures
condition)

\nnext
(3) for any $F$ and $g_{\alpha\beta}$ the conditions $\nabla_\alpha
N^\alpha=0$ and $\nabla_\alpha T^{\alpha\beta}=0$ are satisfied
(conservation conditions)

\nnext
(4) for any $F$ and $g_{\alpha\beta}$ the vector $N^\alpha$ is
future-pointing timelike or zero

\nnext
(5) for any constant $C_1>0$ there exists a positive constant $C_2$
such that for any $F$ and $g_{\alpha\beta}$ with
$-N_\alpha N^\alpha\le C_1$ and any timelike vector $V^\alpha$
the following inequality holds:
$$T^{\alpha\beta}V_\alpha V_\beta\ge C_2(N^\alpha V_\alpha)^2 $$

\nnext
(6) for any constant $C_1>0$ there exists a positive constant $C_2<1$
such that for any $F$ and $g_{\alpha\beta}$ with
$-N^\alpha N_\alpha\le C_1$ and any unit timelike vector $V^\alpha$
$$(g_{\alpha\beta}+V_\alpha V_\beta)T^{\alpha\beta}\le
3C_2 T^{\alpha\beta}V_\alpha V_\beta$$

\nnext
(7) if a solution with the given symmetry of the Einstein equations
coupled to the given matter model is such that the time
coordinate defined above takes all values in the interval
$(t_1,t_2)$ , if it is not possible to extend the spacetime so
as to make this interval longer and if $t_1$ or $t_2$ is finite
then $H(t)$ is unbounded in a neighbourhood of $t_1$ or $t_2$
respectively.

\nnext
(8) for any constant $C_1>0$ there exists a constant $C_2>0$ such
that $T_{\alpha\beta}T^{\alpha\beta}\le C_1$ implies $-N_\alpha
N^\alpha\le C_2$

Some comments will now be made concerning the physical motivation
of some of these conditions. If a given type of matter
can be considered as being made up of particles then a particle
current density $N^\alpha$ is defined. If the particles have
positive rest mass then this vector is future pointing timelike
or zero as required by condition (4). If the particles are
massless then this condition is still satisfied except
for very special types of matter where $N^\alpha$ might be null. If
particles cannot be created or destroyed then $N^\alpha$ is
divergenceless as required in condition (3). It is not easy to
give an intuitive interpretation of conditions (5) and (6). The
meaning of (5) is roughly as follows. If matter is observed
{}from a boosted frame then the particle density is multiplied
by a $\gamma$-factor arising from the effect of Lorentz contraction
on the volume element. The observed energy density is also affected in
this way but picks up an additional $\gamma$-factor. Hence when a
given matter distribution is considered from a boosted frame the
multiplicative factor in the observed energy density behaves like
the square of that in the observed particle density. As for condition
(6), it will be seen in Section 4 that for a perfect fluid it is
related to the condition that the speed of sound should be bounded
away from the speed of light. The given symmetry
referred to in condition (7) will be Bianchi symmetry or
Kantowski-Sachs symmetry, according to the context.

In the following the two fundamental questions formulated above
will be answered for locally spatially homogeneous spacetimes
which are spatially compact and where the matter content satisfies
(some of) the conditions (1)-(8). In fact it will also be necessary
when examining the issue of curvature singularities to assume that
the spacetimes considered are not empty since in the vacuum case
it is much more difficult to separate spacetimes with curvature
singularities from spacetimes with Cauchy horizons. The question
of which spatially compact, locally spatially homogeneous vacuum
spacetimes have Cauchy horizons has been investigated in [3]. In
the following charged matter and field-theoretic matter models (e.g.
Yang-Mills, $\sigma$-models) are not considered. It would
be of interest to know whether similar results hold in those
cases.

The paper is organized as follows. Section 2 contains results on
geodesic completeness in an expanding phase while Section 3 is
concerned with curvature singularities in a contracting phase.
In both cases the hypotheses on the matter model are of a general
nature. Section 4 is concerned with verifying that these properties
hold for various specific matter models, namely perfect fluids,
non-interacting mixtures of perfect fluids and collisionless
matter. As already mentioned, the last section is devoted to
Kantowski-Sachs models.

\vskip 1cm
\noindent
{\bf 2. Geodesic completeness}

In this section it will be shown that under general conditions
there can be no singularity during an expanding phase of a locally
spatially homogeneous spacetime. For this statement the existence
of a {\it compact} Cauchy surface is irrelevant. An expanding
phase means by definition a time interval $(t_1,t_2)$ where $H<0$.
The statement that there are no singularities means that either the
solution can be extended to an interval $(t_1, t_2')$ with $t_2'>
t_2$ or that it is future complete (i.e. that all inextendible
causal geodesics are complete in the future direction). This
statement is a consequence of the following theorem.

\noindent
{\bf Theorem 2.1} Let a spatially locally homogeneous solution of
the Einstein equations coupled to a matter model satisfying the
conditions (1), (2) and (7) be given. Suppose that a Gaussian
time coordinate defined on this spacetime takes all values in
the interval $(t_1,t_2)$ and that the spacetime cannot be extended
so as to enlarge this interval. Suppose further that the mean
curvature $H$ satisfies $\limsup_{t\to t_2} H(t)<\infty$. Then
$t_2=\infty$ and the spacetime is future complete.

\noindent
{\bf Proof} The proof given here is for Bianchi models; for
Kantowski-Sachs models see Section 5. Since spatial compactness
plays no role we may assume without loss of generality that
the spacetime is simply connected. Now some of the Einstein
equations for Bianchi models are needed, namely the Hamiltonian
constraint
$$R-k_{ij}k^{ij}+H^2=16\pi T^{00}\eqno(2.1)$$
and the evolution equation
$$\d_t H=R+H^2-12\pi T^{00}+4\pi g_{ij}T^{ij}.\eqno(2.2)$$
Here $R$ denotes the scalar curvature of the spatial metric.
Combining these two equations gives
$$\d_t H=k_{ij}k^{ij}+4\pi(T^{00}+g_{ij}T^{ij}).\eqno(2.3)$$
Now $k_{ij}k^{ij}\ge\f13 H^2$ and so equation (2.3) together
with conditions (1) and (2) implies that $\d_t H\ge \f13 H^2$.
In particular this shows that $H$ is non-decreasing. Thus,
under the hypotheses of the theorem, the fact that $t_2=\infty$
follows immediately from condition (7). The inequality also
shows that if $H$ is ever positive it must become infinite
in finite time. Since it is already known that $t_2=\infty$
it follows that $H\le 0$. There are now two possibilities:
either $H$ vanishes at some time or it is everywhere negative.

Suppose that $H=0$ at some time $t_0$.
Since $H$ is everywhere non-positive it can be concluded that
$\d_t H$ is also zero there. From equation (2.3) it follows
that $k_{ij}=0$ and $T^{00}=0$ for $t=t_0$. The dominant
energy condition implies that $T^{\alpha\beta}=0$ at $t=t_0$
and hence (see [10], p. 94) everywhere. Thus in this case the
solution must be a vacuum solution. Under these conditions the
Einstein evolution equations imply that the Ricci tensor $R_{ij}$
of $g_{ij}$ is zero at $t_0$. Hence the metric $g_{ij}(t_0)$ is
flat. Uniqueness in the Cauchy problem for the vacuum Einstein
equations now shows that in the first case the spacetime is obtained
by an identification of (the whole of) Minkowski space and
future completeness is clear.

Now consider the second case. There the universe expands for ever.
It has been shown by Lin and Wald [11] that for Bianchi IX models
conditions (1) and (2) imply that the universe cannot expand for
ever. Thus under the hypotheses of the theorem the Bianchi type
cannot be IX. It can then be concluded that $R\le 0$. (For a
simple proof of this see [15].) Using this in (2.1) gives
$$k_{ij}k^{ij}\le H^2\eqno(2.4)$$
The eigenvalues of $k_{ij}$ with respect to $g_{ij}$ are the
solutions of
$$\det (k_{ij}-\lambda g_{ij})=0\eqno(2.5)$$
Call them $\lambda_1$, $\lambda_2$ and $\lambda_3$. Then
$$\lambda_1+\lambda_2+\lambda_3=H,\eqno(2.6)$$
and
$$\lambda_1^2+\lambda_2^2+\lambda_3^2=k_{ij}k^{ij}.\eqno(2.7)$$
Define $p_i=\lambda_i/H$ so that
$$\eqalign{
p_1+p_2+p_3&=1,                 \cr
p_1^2+p_2^2+p_3^2&\le 1.}\eqno(2.8)$$
Let $K$ denote the subset of $\R^3$ where the relations (2.8)
are satisfied. The minimum value of $p_1$ on $K$ will now be
determined. By symmetry this will also be the minimum value
of $p_2$ and $p_3$. The minimum must occur on the boundary
of $K$ and the boundary is given by replacing the inequality
in (2.8) by equality. Since $K$ is convex the minimum must
occur when two of the eigenvalues are equal and using this
its value can be computed to be $-1/3$. It follows that for
any vector $x^i$ tangent to the hypersurfaces of homogeneity
the following inequality holds
$$k_{ij}x^ix^j\le -\f13 Hg_{ij}x^ix^j.\eqno(2.9)$$

Consider now a future directed causal geodesic and let $q^i$ be
the frame components of the projection of its tangent vector
onto the hypersurfaces of constant $t$. Then
$$\eqalign{
{d\over dt}(g_{ij}q^iq^j)&=2k_{ij}q^iq^j     \cr
&\le -\f23 Hg_{ij}q^i q^j}\eqno(2.10)$$
Using the inequality $\d_t H\ge \f13 H^2$ once more shows that
$$H(t)\ge -3/(C+t)\eqno(2.11)$$
for some positive constant $C$. Combining (2.10) and (2.11)
gives
$${d\over dt}(\log g_{ij}q^iq^j)\le 2/(C+t).\eqno(2.12)$$
Hence
$$(g_{ij}q^iq^j)^{-1/2}\ge C(1+t)^{-1}.\eqno(2.13)$$
The affine parameter length of the geodesic up to a certain point
is given by the integral of $(\epsilon+g_{ij}q^iq^j)^{-1/2}$ with
respect to $t$, where $\epsilon$ is $1$ for a timelike geodesic
parametrized by arc length and zero for a null geodesic. Future
geodesic completeness is equivalent to the divergence of
this integral as the upper limit tends to infinity. This is
guaranteed by (2.13).

\vskip 1cm
\noindent
{\bf 3. Curvature singularities}

The subject of this section is a contracting phase of the universe
i.e. a time interval $(t_1,t_2)$ where $H>0$. Only Bianchi models are
considered here - analogous results for spacetimes with
Kantowski-Sachs symmetry are presented in Section 5.  It was shown in
the last section that conditions (1) and (2) imply that $H$ is
non-decreasing. It follows that any extension of the solution to an
interval $(t_1,t_2')$ with $t_2'>t_2$ must also be a contracting phase
in this sense. Extend the time interval as far as possible to the
future and call the result $(t_1,t_*)$. It was also remarked in the
last section that if $H$ is positive somewhere it must become infinite
in finite time. Hence $t_*<\infty$. In this section, in contrast to
the previous section, the spatial compactness of the model plays an
important role. Also all the conditions (1)-(7) will be used. Writing
out the equation $\nabla_\alpha N^\alpha=0$ explicitly for a Bianchi
model gives
$$\d_t((\det g)^{1/2} N^0)=-C^i_{ij}N^j(\det g)^{1/2}\eqno(3.1)$$
Here $C^i_{jk}$ are the structure constants of the Lie algebra
corresponding to the Lie group being considered. This equation
should be thought of as being defined on the universal covering
spacetime where an invariant frame exists. In the Bianchi types
of class A where, by definition, $C^i_{ij}=0$ the right hand
side of (3.1) vanishes identically and the quantity
$(\det g)^{1/2}N^0$ is conserved. It will now be shown that
the assumption of spatial compactness is enough to guarantee
the vanishing of the right hand side of (3.1), even for class B
models. To see this let $X$ be any vector on the Lie group
$G$ which is left invariant and which projects to a smooth
vector field $\bar X$ on $M$. The divergence of $X$ is constant
because it is invariant and the divergence of $\bar X$ is the
same constant. However the integral of ${\rm div}\ \bar X$ over
$M$ must be zero by Stokes' theorem. It follows that $X$ has
divergence zero. In terms of components this divergence is just
$C^i_{ij}X^j$. Now $N^j$ is at each time a vector field whose
pull-back to the universal covering manifold has the properties
of the vector field $X$ just discussed. It follows that the right
hand side of (3.1) is zero as claimed.

The aim now is to show that if conditions (1)-(7) hold and if
$N^\alpha$ does not vanish then the singularity which occurs
as $t\to t_*$ is a curvature singularity. Because of the
conserved quantity which has been found it suffices to assume
that $N^\alpha$ does not vanish on the initial hypersurface
to ensure that it does not vanish anywhere.

\noindent
{\bf Theorem 3.1} Let a solution with Bianchi
symmetry of the Einstein equations coupled to a matter model
satisfying conditions (1)-(7) and admitting a compact Cauchy
hypersurface be given. Suppose that a Gaussian time coordinate defined
on this spacetime takes all values in the interval $(t_1,t_*)$ and
that the spacetime cannot be extended so as to enlarge this
interval. Suppose further that $H$ is positive somewhere and that the
particle current density $N^\alpha$ does not vanish. Then the mass
density measured by any observer is unbounded as $t\to
t_*$. If in addition condition (8) is satisfied then $\limsup_{t\to
t_*}G^{\alpha\beta} G_{\alpha\beta}(t)=\infty$, where
$G_{\alpha\beta}$ is the Einstein tensor of $g_{\alpha\beta}$.

\noindent
{\bf Proof} The first thing to be done is to obtain some control
on the rate at which $H(t)$ blows up as $t\to t_*$. (That it does
blow up as $t\to t_*$ follows from condition (7).) It was already
mentioned in the previous section that $\d_t H\ge\f13 H^2$.
Integrating this inequality gives
$$H(t)\le{3\over t_*-t}\eqno(3.2)$$
On the other hand (2.2) and the dominant energy condition imply
that
$$\d_t H\le R+H^2\eqno(3.3)$$
Let $R_+(t)=\max(R(t),0)$. Then $\d_t H\le R_++H^2$. For $t$ close
to $t_*$ we have $H\ge 1$. Thus $\d_t H\le R_+H+H^2$ and integrating
this inequality gives the estimate
$$H(t)\ge(t_*-t)^{-1}\exp\left(-\int^{t_*}_t R_+(t')dt'\right)
\eqno(3.4)$$
Here the right hand side is defined to be zero if the integral
is infinite.
If the Bianchi type is not IX then $R_+=0$ and (3.4) simplifies to
$$H(t)\ge(t_*-t)^{-1}\eqno(3.5)$$
{}From these estimates for $H$ it is possible to get estimates for
$\det g$ since $d/dt(\log\det g)=-2H$. It follows from (3.2) that
$$\det g\ge C_1(t_*-t)^{6}\eqno(3.6)$$
for some positive constant $C_1$. If $\int_t^{t_*}R_+(t)dt<\infty$
for any $t<t_*$ then (3.4) implies that
$$\det g(t)\le C_2(t_*-t)^2\eqno(3.7)$$
The integrability condition on the scalar curvature is obviously
satisfied when the Bianchi type is not IX.  In type IX a more
detailed analysis is necessary to see whether it holds. When (3.7)
holds the fact that $N^0(\det g)^{1/2}$ is conserved shows that
$$N^0\ge C_3(t_*-t)^{-1}\eqno(3.8)$$
Applying condition (5) to the unit normal vector to the
homogeneous hypersurfaces gives
$$T^{00}\ge C N_0^2\eqno(3.9)$$
as long as $N_\alpha N^\alpha$ is bounded.

Suppose now that the Bianchi type is not IX. It will be shown
that the assumption that $N_\alpha N^\alpha$ is bounded leads
to a contradiction in this case. If this quantity is bounded
then (3.8) and (3.9) imply that there exists a constant $C>0$
such that
$$T^{00}\ge C(t_*-t)^{-2}\eqno(3.10)$$
Then the Hamiltonian constraint (2.1) and the fact that $R\le0$
when the Bianchi type is not IX give an estimate of the form
$$k_{ij}k^{ij}\le H^2(1-\eta),\ \eta>0\eqno(3.11)$$
Taking a linear combination of (2.1) and (2.2) gives
$$\d_t H=\alpha(k_{ij}k^{ij}-H^2)+(1-\alpha)R+H^2-4\pi(3-4\alpha)
T^{00}+4\pi g_{ij}T^{ij}\eqno(3.12)$$
Because of (6) it is possible to choose $\alpha>0$ so that the
sum of the last two terms in this inequality is not positive.
Hence
$$\d_t H\le (1-\alpha\eta)H^2\eqno(3.13)$$
This allows the inequality (3.8) to be improved to
$$N^0\ge C(t_*-t)^{{-1\over 1-\alpha\eta}}\eqno(3.14)$$
{}From (3.9) it follows that
$$T^{00}\ge C(t_*-t)^{{-2\over 1-\alpha\eta}}\eqno(3.15)$$
This contradicts the Hamiltonian constraint. The conclusion is that
if the Bianchi type is not IX then $N^\alpha N_\alpha$ must be
unbounded. Consider now any observer, represented
mathematically by a timelike curve with unit tangent vector $t^\alpha$.
Then the mass density measured by this observer is
$$-N^\alpha t_\alpha\ge\sqrt{-N_\alpha N^\alpha}\eqno(3.16)$$
This gives the first conclusion of the theorem in the case that
the Bianchi type is not IX. The other conclusion follows
immediately from the fact that $N^\alpha N_\alpha$ is unbounded.

Consider now the Bianchi IX case. It is a standard fact (see e.g.
[15]) that the
dimensionless quantity $(\det g)^{1/3}R$ is bounded above. Hence
$R\le C(\det g)^{-1/3}$. Suppose for a moment that
$$\det g\ge C(t_*-t)\eqno(3.17)$$
for some positive constant $C$. Then $R_+\le C(t_*-t)^{-1/3}$.
This implies the integrability of $R_+$ and hence, by (3.4),
that $H(t)\ge C/(t_*-t)$. It follows that $\det g$ tends to zero
as $t\to t_*$. If, on the other hand, the inequality (3.17) does
not hold for any constant $C$ then there must exist a sequence
of times $t_n$ with $t_n\to t_*$ such that
$$\det g(t_n)\le C(t_*-t_n)\eqno(3.18)$$
Hence, in particular, $\det g(t_n)\to 0$. Since $\det g$ is
monotonically decreasing near the singularity it follows
that $\det g\to0$ in this case too.

It will now be shown that the assumption that $N_{\alpha}N^\alpha$
is bounded leads to a contradiction. This suffices to complete the
proof of the theorem since from that point on the argument is
identical to that given above for the case where the Bianchi type is
not IX. In the equation (2.2) the combination $R-12\pi T^{00}
+4\pi g_{ij}T^{ij}$ occurs. If $N^\alpha N_\alpha$ is bounded
then condition (6) allows the sum of the second and third terms to be
bounded from above by $-CT^{00}$ for some positive constant $C$.
Also condition (5) gives
$$T^{00}\ge C(N^0)^2=C'(\det g)^{-1}\eqno(3.19)$$
Thus as $\det g$ tends to zero $CT^{00}$ becomes larger
than $R$ and the above combination of terms from (2.2) is negative.
Hence $\d_t H\le H^2$. It follows that the inequalities (3.7),
(3.8) and (3.10) remain true in the Bianchi IX case. The
inequality (3.11) can also be obtained if we replace the fact that
$R\le 0$ used before by the fact that sufficiently near the
singularity $R\le CT^{00}$ for any given positive constant $C$.
This can also be used to control the term involving $R$ in (3.12)
using a small constant multiplied by $T^{00}$. This is enough to
prove (3.13)-(3.15) in the Bianchi IX case. The same trick of
dominating $R$ by a small amount of $T^{00}$ shows that (3.15)
contradicts the Hamiltonian constraint in this case also and this
completes the proof of the theorem.

\noindent
{\bf Remarks} 1. When applied to a general perfect fluid solution
of Bianchi type IX this theorem is a non-trivial generalization of
the results given in [6].

\noindent
2. In [12] it was claimed that a singularity in a
contracting phase of a Bianchi class A non-vacuum solution of the
Vlasov-Einstein system must be a curvature singularity. The claim
is not justified by the arguments given there but it is true.
It follows from Theorem 3.1 above and the fact, proved in the
next section, that the Vlasov equation satisfies conditions (1)-(8).

\noindent
3. For solutions where $N^\alpha$ is orthogonal to the hypersurfaces
$t=$const. $\sqrt{-N^\alpha N_\alpha}=N^0$ and so in that case the
theorem remains true if conditions (5) and  (6) are dropped from the
assumptions.

\vskip 1cm
\noindent
{\bf 4. Matter models}

Before specific matter models are studied some general remarks
will be made concerning condition (7) in spacetimes with Bianchi
symmetry. Suppose that a solution
is given on an interval $(t_1,t_2)$, that $t_2$ is finite and
that $H$ is bounded on a neighbourhood of $t_2$. Then $\det g$
and its inverse are bounded on a neighbourhood of $t_2$. It
is a standard fact for Bianchi models that under these conditions
the scalar curvature $R$ is bounded above (see e. g. [15]).
Next the Hamiltonian constraint shows that $k^{ij}k_{ij}$ is bounded.
An argument given in [12] (p. 88-89) then shows that $g_{ij}$
and $k_{ij}$ are bounded in a neighbourhood of $t_2$.
Analogous statements hold in the case that the roles of $t_1$ and
$t_2$ are interchanged. What is needed to check condition (7) is that
when a solution of the Einstein-matter equations is given on
some open interval and when $g_{ij}$, $k_{ij}$ and $(\det g)^{-1}$
are bounded near an endpoint of this interval, then the solution
extends through this endpoint to a strictly longer interval.
Call this condition (7$'$).

It is now time to look at some examples. The matter model which
has been studied most extensively in the context of Bianchi
models is the perfect fluid. It will now be investigated under
which conditions on the equation of state a perfect fluid satisfies
conditions (1)-(8). The equation of state is a relation $p=f(\rho)$
between energy density and pressure. The energy-momentum tensor is
$T^{\alpha\beta}=\rho u^\alpha u^\beta+(\rho+p)g^{\alpha\beta}$
The equation of state will be assumed to satisfy the following
conditions

\next
(i) $f$ is a continuous function from $[0,\infty)$ to the non-negative
real numbers with $f(0)=0$ which is $C^1$ for $\rho>0$

\next
(ii) $0\le df/d\rho\le 1$ for each $\rho>0$

Condition (1) is satisfied if $|p|\le\rho$ ([10], p. 91) while
condition (2) is equivalent to $p\ge0$. Hence both of these
conditions are consequences of (i) and (ii). The quantity
$N^\alpha$ is defined to be $ru^\alpha$, where
$$r=\exp\left\{\int_1^\rho(\rho'+f(\rho'))^{-1}d\rho'\right\}\eqno(4.1)$$
This definition ensures that the Euler equations
$\nabla_\alpha T^{\alpha\beta}=0$ imply $\nabla_\alpha N^\alpha=0$
and hence that (3) is satisfied. Condition (4) obviously holds. A
sufficient condition for (5) to be true is that an upper bound on
$r$ implies an inequality of the form $\rho\ge Cr^2$. In fact this
follows from (4.1) and the dominant energy condition. Condition (6)
represents a restriction on the equation of state which goes
beyond (i) and (ii), namely that when $\rho$ is less than some
given constant $C_1$ there is a constant $C_2<1$ such that
$p\le C_2\rho$. It will be expressed by saying that the
fluid is not asymptotically stiff at low densities. Condition
(8) obviously holds.

The one condition which remains to be verified for a perfect
fluid is (7$'$). In order to do this it is useful to write
the Euler equation $\nabla_\alpha T^{\alpha\beta}=0$ in terms
of the density $\rho$ and the spatial components $v^i$ of the
four-velocity (expressed in an invariant frame). Then
no spatial derivatives occur and a system of ordinary differential
equations is obtained. If we denote the matter variables $\rho$ and
$u^i$ collectively by $f$ then the equations take the form
$$A(f,g_{ij})df/dt=F(f,g_{ij}, k_{ij}),\eqno(4.2)$$
where the matrix $A$ and the vector $F$ depend smoothly on
their arguments. As long as $\rho>0$ the determinant of $A$ is
non-vanishing. Hence the evolution equation can be written as
$df/dt=A^{-1}F$. The right hand side of this latter equation is
regular so long as $\det g$ and $\rho$ are bounded away from zero.
This equation can be combined with the evolution equations for the
geometry, which in the homogeneous case are
$$\eqalignno{\d_t g_{ij}&=-2k_{ij}&(4.3)                  \cr
\d_t k_{ij}&=R_{ij}+(g^{lm}k_{lm})k_{ij}-2k_{il}k^l_j-8\pi T_{ij}
-4\pi T^{00}g_{ij}+4\pi(g_{lm}T^{lm})g_{ij}&(4.4)}$$
The Ricci tensor $R_{ij}$ is a rational function of $g_{ij}$ whose
exact form depends on the Bianchi type. Taking the equation for
$df/dt$ together with (4.3) and (4.4) gives a system of ordinary
differential equations for $f$, $g_{ij}$ and $k_{ij}$. The
coefficients in these equations are regular as long as $\det g$ and
$\rho$ are strictly positive. Standard results on existence and
uniqueness for ordinary differential equations (see e.g. [9])
show that given initial data $(\rho, u^i, g_{ij}, k_{ij})$ with
$\rho>0$ there exists a unique solution of these equations on
some time interval with the given initial values. Moreover this
solution can be extended as long as $u^i,g_{ij}, k_{ij},\rho,
\rho^{-1}$ and $(\det g)^{-1}$ remain bounded.

To verify condition (7$'$) it remains to show that if a solution
of the Einstein-Euler system is given on some time interval
and if $g_{ij}$, $k_{ij}$ and $(\det g)^{-1}$ are bounded then the
relevant matter quantities are bounded. The conservation law
for $N^0$ shows that $ru^0$ and its inverse are bounded.
Hence it can be concluded that $r$ is bounded above.
For a perfect fluid condition (6) implies that for $r\le C_1$ an
inequality of the form $\rho\ge C_2 r^{2-\epsilon}$ holds, for
some $\epsilon>0$. Now
$$\eqalign{
T^{00}&\ge\rho (u^0)^2\ge C r^{2-\epsilon}(u^0)^2    \cr
      &=C(ru^0)^2r^{-\epsilon}}\eqno(4.5)$$
The Hamiltonian constraint shows that $T^{00}$ is bounded and
so it follows that $r^{-1}$ is bounded. Combining this with
the boundedness of $ru^0$ shows that $u^0$ is bounded. This
completes the verification of condition (7$'$) for a perfect
fluid.

The above information concerning perfect fluids will now be
collected.

\noindent
{\bf Proposition 4.1} Consider a perfect fluid with equation of
state $p=f(\rho)$ satisfying conditions (i) and (ii). Then
conditions (1)-(5)and (8) are satisfied. If, in addition, the
fluid is not asymptotically stiff at low densities then conditions
(6) and (7) are also satisfied.

\vskip 10pt\noindent
It follows from this proposition that Theorem 3.1 applies to
any perfect fluid which satisfies the conditions (i) and (ii)
and which is not asymptotically stiff at low densities.

Another kind of matter model which has been popular in cosmology
is that of $n$ (usually two) non-interacting perfect fluids
[4]. The energy-momentum tensor in this case is
of the form
$$T^{\alpha\beta}=T^{\alpha\beta}_{(1)}+\ldots+T^{\alpha\beta}_{(n)}
\eqno(4.6)$$
where each summand $T^{\alpha\beta}_{(i)}$ has the form of the
energy-momentum tensor of a perfect fluid with its own four-velocity
$u^\alpha_{(i)}$, density $\rho_{(i)}$ and pressure $p_{(i)}$. The
latter are related by equations of state
$p_{(i)}=f_{(i)}(\rho_{(i)})$.
The condition that the fluids are non-interacting is expressed by
the equations $\nabla_\alpha T^{\alpha\beta}_{(i)}=0$ for each $i$.
If we define $r_{(i)}$ by putting an index $i$ on all the quantities
in (4.1) and let $N^\alpha_{(i)}=r_{(i)} u^\alpha_{(i)}$ then it
follows that $\nabla_\alpha N^\alpha_{(i)}=0$ for each $i$. In
particular the sum $N^\alpha$ of the $N^\alpha_{(i)}$ has zero
divergence. Before showing under which assumptions a mixture of
perfect fluids satisfies conditions (1)-(8), a more general result
will be given on matter models which are composites of two others.

\noindent
{\bf Lemma 4.1} Let two matter models be given with matter variables
$F_{(1)}$ and $F_{(2)}$, particle current densities $N_{(1)}^\alpha$
and $N_{(2)}^\alpha$ and energy-momentum tensors $T_{(1)}^{\alpha
\beta}$ and $T_{(2)}^{\alpha\beta}$. Define a new (composite)
matter model by taking as matter variables $F=(F_{(1)}, F_{(2)})$
and defining $N^\alpha$ and $T^{\alpha\beta}$ as the sums of the
corresponding quantities for the individual matter models. If the
original matter models satisfy conditions (1)-(6) then the
composite model does so too.

\noindent
{\bf Proof} It is obvious that the composite matter model satisfies
conditions (1)-(4). Next note that $-N^\alpha N_\alpha\ge
-N^\alpha_{(1)}N_\alpha^{(1)}-N^\alpha_{(2)}N_\alpha^{(2)}$, so
that the boundedness of $N^\alpha N_\alpha$ implies that of
$N^\alpha_{(i)} N_\alpha^{(i)}$ for $i=1,2$. It follows that
condition (6) is satisfied and condition (5) is a consequence
of the following computation:
$$\eqalign{
T_{\alpha\beta}V^\alpha
V^\beta&=T_{\alpha\beta}^{(1)}V^\alpha V^\beta
+T_{\alpha\beta}^{(2)}V^\alpha V^\beta       \cr
&\ge C_2 [(N_\alpha^{(1)}V^\alpha)^2
+(N_\alpha^{(2)}V^\alpha)^2]            \cr
&\ge \f12 C_2(N^\alpha V_\alpha)^2}
\eqno(4.7)$$

\vskip 10pt\noindent
Consider now a non-interacting mixture of perfect fluids and suppose
that each individual fluid satisfies the
hypotheses of Proposition 4.1 including the condition that it
should not be asymptotically stiff. It then follows from Lemma 4.1
that the mixture satisfies conditions (1)-(6). It will now be investigated
whether conditions (7) and (8) also hold. The evolution equations for a
mixture of non-interacting fluids can be written in a similar way to those
for a single fluid. Thus to verify condition (7) in this case it
suffices to check that the fluid quantities remain finite in a
given regular spacetime. However each fluid satisfies the Euler
equations separately and so the argument given above for a single
fluid applies without change. (It must, of course, be assumed for
this that none of the fluids are asymptotically stiff at low
densities.) A straightforward calculation shows that
$$T^{\alpha\beta}T_{\alpha\beta}\ge (\sum_{(i)}\rho_{(i)})^2\eqno(4.8)$$
Hence if $T^{\alpha\beta}T_{\alpha\beta}$ is bounded all the
$\rho_{(i)}$ are bounded. Moreover, the boundedness of
$T^{\alpha\beta}T_{\alpha\beta}$ implies that the quantities
$\rho_{(i)}\rho_{(j)}(u^\alpha_{(i)}u_\alpha^{(j)})^2$ are bounded
for each $i,j$. It follows that $N^\alpha_{(i)}N_\alpha^{(j)}$ is
bounded for each $i,j$. Thus condition (8) is also satisfied and the
following analogue of Proposition 4.1 is obtained.

\noindent
{\bf Proposition 4.2} Consider a mixture of non-interacting perfect
fluids, each of which satisfies the hypotheses of Proposition 4.1
including the condition that none of them be asymptotically stiff
at low densities. Then conditions (1)-(8) are satisfied.

The next matter model which will be considered is a collisionless
gas described by the Vlasov equation. For an introduction to the
Vlasov-Einstein system see [7]. Some general information on
spatially homogeneous solutions can be found in [12]. In the
following we consider a gas of particles of unit mass. The
distribution function $f$ is supposed to be a $C^1$ function
with compact support in velocity space initially and these
properties are preserved by the time evolution. The
quantities $N^\alpha$ and $T^{\alpha\beta}$ are defined in
the usual way as certain weighted integrals of
$f$. Conditions (1)-(4) are standard properties of
solutions of the Vlasov-Einstein system. Condition (7) for this
matter model is the statement of Lemma 2.2 of [12]. It remains
to check conditions (5), (6) and (8) and to do this we may
choose a frame whose timelike member is $V^\alpha$ in order
to do the calculation. Then the inequalities of (5) and (6) become
$\hat T^{00}\ge C_2(\hat N^0)^2$ and $\delta^{ij}\hat T_{ij}\le 3C_2
\hat T^{00}$. (The hats here indicate the use of indices associated
to an orthonormal frame.)
$$\eqalign{
-N_\alpha N^\alpha&=-\left(\int f(p^a)p_\alpha/p^0 dp^1 dp^2 dp^3
                     \right)
                     \left(\int f(q^a)q^\alpha/q^0 dq^1 dq^2 dq^3
                     \right)   \cr
&=-\int\int f(p^a)f(q^a)p^\alpha q_\alpha  /(p^0q^0)dp^1 dp^2 dp^3
dq^1 dq^2 dq^3
\cr
&\ge\int\int f(p^a)f(q^a)/(p^0q^0) dp^1 dp^2 dp^3 dq^1 dq^2 dq^3    \cr
&=\left(\int f(p^a)/p^0 dp^1 dp^2 dp^3\right)^2}\eqno(4.9)$$
Hence
$$\eqalign{
N^0&=\int f(p^a) dp^1 dp^2 dp^3                  \cr
   &\le \left(\int f(p^a) p^0 dp^1 dp^2 dp^3\right)^{1/2}
      \left(\int f(p^a)/p^0
   dp^1 dp^2 dp^3\right)^{1/2}                    \cr
   &\le (T^{00})^{1/2}(-N_\alpha N^\alpha)^{1/4}}\eqno(4.10)$$
This shows that (5) holds. It follows directly from the definitions
that the inequality of (6) holds with $C_2=1/3$ even without
restricting $N_\alpha N^\alpha$ to be bounded. Finally
$$\eqalign{
T_{\alpha\beta}T^{\alpha\beta}&=\left(\int f(p^a)p_\alpha p_\beta/p^0
dp^1 dp^2 dp^3\right)\left(\int f(q^a)q^\alpha q^\beta/q^0
dq^1 dq^2 dq^3\right)             \cr
&=\int\int f(p^a)f(q^a)(p^\alpha q_\alpha)^2/(p^0q^0) dp^1 dp^2 dp^3
dq^1 dq^2 dq^3                   \cr
&\ge -\int\int f(p^a)f(q^a)(p^\alpha q_\alpha)/(p^0q^0) dp^1 dp^2 dp^3
dq^1 dq^2 dq^3                   \cr
&=-N_\alpha N^\alpha}\eqno(4.11)$$

\noindent
{\bf Proposition 4.3} Consider a collisionless gas of particles of
unit mass for which the distribution function is $C^1$ and compactly
supported in velocity space. Then conditions (1)-(8) are satisfied.

\vskip 10pt
Most of what was said in the last paragraph only used the fact that
matter is described by a distribution function and the expression
for the energy-momentum tensor in terms of this distribution
function. Hence it has more general applicability to matter
models based on kinetic theory. Consider the case of the Boltzmann
equation. It follows from the above that conditions (1)-(6) and
(8) are satisfied for the Boltzmann equation. The one condition
which causes difficulty is (7). Recall furthermore that to verify
(7) it is enough to verify (7$'$). For the Boltzmann equation this
is a non-trivial task. Even in the case of spatially homogeneous
solutions of the classical Boltzmann equation it is difficult to
prove global existence and uniqueness theorems. (For details see
[2] and references therein.) Just how difficult it is depends on
the form of the collision kernel. For the special relativistic
Boltzmann equation there appear to be no results in the literature
on global existence of unique solutions for general homogeneous
initial data. (There is a result for inhomogeneous data close to
equilibrium [8].)

\vskip 1cm
\noindent
{\bf 5. Kantowski-Sachs spacetimes}

In this section a contracting phase of the universe will be
considered as in Section 3 but now the symmetry is of
Kantowski-Sachs type rather than Bianchi type. It should
be mentioned that for this symmetry type no singularities
occur in an expanding phase if conditions (1), (2) and (7)
hold. It can be shown as in the Bianchi case that if (7)
holds either the solution can be extended to the future beyond
the expanding phase or the solution exists globally in a
Gaussian time coordinate and expands for ever. The
impossibility of the second alternative under conditions
(1) and (2) has been shown by Burnett[1].

Suppose now that a solution is defined on an interval $(t_1,t_*)$
which has been maximally extended towards the future. Suppose
further that this is a contracting phase.
In a Kantowski-Sachs spacetime the metric can be written in the
form
$$ds^2=-dt^2+A^2(t)dx^2+B^2(t)d\Sigma^2\eqno(5.1)$$
where $d\Sigma^2$ is the standard metric on the two-sphere.
It is easy to check that the divergence of the second fundamental
form of the hypersurfaces $t$=const. is zero. It follows from
the momentum constraint that $\d/\d t$ is an eigenvector of
the energy-momentum tensor with eigenvalue $T^{00}$. Hence
$$T^{\alpha\beta}T_{\alpha\beta}\ge (T^{00})^2\eqno(5.2)$$
It can be shown that, as in the case of the Bianchi models,
$(\det g)^{1/2}N^0$ is time independent. In this case $\det g$
denotes the determinant of the three-dimensional metric in
the above coordinates, so that $(\det g)^{1/2}=AB^2$. The
following theorem is an analogue of Theorem 3.1 for spacetimes
with Kantowski-Sachs symmetry.

\noindent
{\bf Theorem 5.1} Let a solution with Kantowski-Sachs
symmetry of the Einstein equations coupled to a matter model
satisfying conditions (1)-(7) and admitting a compact Cauchy
hypersurface be given. Suppose that a Gaussian time coordinate defined
on this spacetime takes all values in the interval $(t_1,t_*)$ and
that the spacetime cannot be extended so as to enlarge this
interval. Suppose further that $H$ is positive somewhere and that the
particle current density $N^\alpha$ does not vanish. Then $t=t_*$ is
a curvature singularity in the sense that some curvature invariant
blows up uniformly as $t\to t_*$.

\noindent
{\bf Proof} Note first that, as a consequence of condition (7)
$H$ tends to infinity as $t\to t_*$.
A spacetime with Kantowski-Sachs symmetry is, in
particular, spherically symmetric. It is thus possible to define
the area radius $r$ which, in terms of the coordinates used above, is
just $B$. The gradient of $r$ is everywhere timelike and its length
squared is $-\dot B^2$. The mass function $m$ is defined in a
spherically symmetric spacetime by
$$1-2m/r=\nabla_\alpha r\nabla^\alpha r\eqno(5.3)$$
Thus in the Kantowski-Sachs case
$$m=\f12 B(1+\dot B)^2\eqno(5.4)$$
In a spherically symmetric spacetime $m$ is monotone increasing
along the integral curves of $\nabla^a r$ as long as the latter
are timelike[1]. Hence $m$ can be bounded from
below in terms of a positive constant only depending on the
initial data. When studying spherically symmetric spacetimes
it is useful to consider quantities defined on the quotient
of spacetime by the group action. This space inherits a
2-dimensional metric $g_{ab}$ with Gaussian curvature $K$
and a 2-dimensional energy-momentum tensor $T_{ab}$. Let
$\tr T=g^{ab}T_{ab}$. The Kretschmann scalar is given by
$$\eqalign{
R^{\alpha\beta\gamma\delta}R_{\alpha\beta\gamma\delta}
&=4K^2+4r^{-4}(\nabla^a r\nabla_a r)^2
+r^{-4}\left({1\over 2r}\nabla^a r
\nabla_a r-2\pi r\tr T\right)^2          \cr
&+16\pi^2r^{-2}\left(T_{ab}-{1\over 2}\tr Tg_{ab}\right)
         \left(T^{ab}-{1\over 2}\tr Tg^{ab}\right)}
\eqno(5.5)$$
This formula also applies to spacetimes with plane symmetry and
was used in [13] in that context. All terms except the last are
manifestly non-negative. The last is also non-negative provided
the energy-momentum tensor is diagonalizable. For Kantowski-Sachs
symmetry this is automatic. Hence the Kretschmann scalar can be
bounded below by $16m^2/r^6$. Since it is known that $m$ is
bounded below by a positive constant it follows that if
$\lim_{t\to t_*}r=0$ the Kretschmann scalar will blow up as
$t\to t_*$. Thus to prove the theorem we can assume without
loss of generality that $r$ does not tend to zero at the
singularity. The scalar curvature $R$ of the $t$=const.
hypersurfaces is simply $B^{-2}$. Thus under the present
assumptions $R$ is bounded above. Given this information (3.4) can
be used as in the Bianchi case (when the type is not IX) to show
that $\det g\to 0$ as $t\to t_*$. By the conservation law
$N^0$ tends to infinity in this limit. Condition (5),
applied to the normal vector, shows that $T^{00}$ tends
to infinity in this limit. In the Kantowski-Sachs case
$T^{00}$ is an eigenvalue of the energy-momentum tensor
and so $T^{\alpha\beta}T_{\alpha\beta}\to\infty$ as
$t\to t_*$. This completes the proof.

\vskip 10pt
It may be seen as unsatisfactory that this argument leaves open the
nature of the curvature singularity in any given case. All that the
above proof shows is that either
$R^{\alpha\beta\gamma\delta}R_{\alpha\beta\gamma\delta}$ or
$G^{\alpha\beta}G_{\alpha\beta}$ blows up as the singularity is
approached. It will now be shown with the help of arguments of
Collins[5] that the second of these quantities always blows
up. According to the above proof, if $B$ does not tend to zero as
$t\to t_*$ then the statement is true. One of the field equations
reads
$$2B\ddot B+1+\dot B^2=-T^1_1B^2\eqno(5.6)$$
If $B$ tends to zero as $t\to t_*$ then $\dot B$ must be negative
somewhere. Equation (5.6) then implies that it must stay negative.
It follows from (5.6) that $d/dt(B(1+\dot B^2))\ge 0$. In fact this
is just the monotone property of the mass function mentioned earlier.
It follows that, fixing some time $t_0$ and letting $B_0=B(t_0)$:
$$\dot B(t)^2\ge -1+B(t_0)(1+\dot B(t_0)^2)/B(t)\eqno(5.7)$$
for any $t>t_0$. Since $B\to 0$ it follows that $\dot B\to\infty$.
Hence in particular for any $\epsilon>0$ the inequality
$\epsilon (\dot B/B)^2\ge 1/B^2$ holds close to the
singularity. The explicit form of the Hamiltonian constraint in this
case is
$${2\dot A\dot B\over
AB}+{\dot B^2\over B^2}+{1\over B^2} =8\pi T^{00}\eqno(5.8)$$
Hence
$$H=-\left({\dot A\over A}+{2\dot B\over B}\right)
=\left(8\pi T^{00}-{1\over B^2}
+{3\dot B^2\over B^2}\right)\bigg/\left({-2\dot B\over B}
\right)\eqno(5.9)$$
This implies an inequality of the form
$$H\ge C|\dot B/B|\eqno(5.10)$$
In other words $-d/dt(\log\det g)\ge -Cd/dt(\log B)$ or
$d/dt(\det g/B^C)\le 0$. It follows that $\det g$ tends to
zero as $t\to t_*$. Hence $T^{00}$ is unbounded and the same is
true of $G^{\alpha\beta}G_{\alpha\beta}$.

In order to be able to apply Theorem 5.1 to the matter models
discussed in Section 4, it is necessary to verify that
condition (7) holds for these matter models in the case
of Kantowski-Sachs symmetry. Note first that (7) follows
{}from condition (7$'$) as in the Bianchi case. The proof is
similar to that given for Bianchi symmetry at the beginning
of Section 4. In the Kantowski-Sachs case it is not known
a priori that the scalar curvature $R$ is bounded above.
However the evolution equation (2.3) for the mean curvature
shows that on an interval where $H$ is bounded the integral
in time of $k_{ij}k^{ij}$ is bounded. This is enough to
apply the argument of [12] to obtain the boundedness of
the metric. This, together with the boundedness of $(\det g)^{-1}$,
shows that $R$ is in fact bounded and from that point on the
argument proceeds as before. In the case of collisionless
matter condition (7) was verified in [12]. For a perfect fluid
of mixture of perfect fluids (7$'$) can be checked as much
as in the Bianchi case. The only difference is in the parametrization
of the 4-velocity of each fluid. This necesarily lies in the
$(t,x)$ plane and so can be parametrized by its component in
the $x$-direction.

\vskip 1cm
\noindent
{\bf Acknowledgements} I thank Piotr Chru\'sciel for constructive
criticism of a previous version of this paper.

\vskip 1cm
\noindent
{\bf References}

\noindent
[1] G. BURNETT. Incompleteness theorems for the spherically
symmetric spacetimes. {\it Phys. Rev.} {\bf D43} (1991), 1143-1149.
\next
[2] C. CERCIGNANI. {\it The Boltzmann equation and its applications.}
(Springer, 1988).
\next
[3] P. T. CHRU\'SCIEL, A. D. RENDALL. Unpublished.
\next
[4] A. A. COLEY, J. WAINWRIGHT. Qualitative analysis of two-fluid Bianchi
cosmologies. {\it Class. Quantum Grav.} {\bf 9} (1992), 651-665.
\next
[5] C. B. COLLINS. Global structure of the Kantowski-Sachs
cosmological models. {\it J. Math. Phys.} {\bf 18} (1977), 2116-2124.
\next
[6] C. B. COLLINS, G. F. R. ELLIS. Singularities in
Bianchi cosmologies. {\it Physics Reports} {\bf 56} (1979), 65-105.
\next
[7] J. EHLERS. Survey of general relativity. In {\it Relativity,
Astrophysics and Cosmology.} (Reidel, 1973).
\next
[8] R. GLASSEY, W. STRAUSS. Asymptotic stability of the relativistic
Maxwellian. {\it Publ. R.I.M.S. Kyoto Univ.} {\bf 29} (1993),
301-347.
\next
[9] P. HARTMAN. {\it Ordinary differential equations.} 2nd Ed.
(Birkh\"auser, 1982).
\next
[10] S. W. HAWKING, G. F. R. ELLIS. {\it The large scale structure
of space-time.} (Cambridge University Press, 1973).
\next
[11] X.-F. LIN, R. WALD. Proof of the closed universe
recollapse conjecture for general Bianchi type IX cosmologies.
{\it Phys. Rev.} {\bf D41} (1990), 2444-2448.
\next
[12] A. D. RENDALL. Cosmic censorship for some spatially homogeneous
cosmological models. {\it Ann. Phys.} {\bf 233} (1994), 82-96.
\next
[13] A. D. RENDALL. On the nature of singularities in plane
symmetric scalar field cosmologies. Preprint gr-qc/9408001 (1994) .
\next
[14] I. M. SINGER. Infinitesimally homogeneous spaces. {\it Commun.
Pure Appl. Math.} {\bf 13} (1960), 685-697.
\next
[15] R. WALD. Asymptotic behaviour of homogeneous cosmological
models in the presence of a positive cosmological constant.
{\it Phys. Rev.} {\bf D28} (1983), 2118-2120.
\next
[16] R. WALD. {\it General Relativity.} (University of Chicago
Press, 1984).

\end